\documentstyle[epsfig,12pt]{article}  
\newcommand{\bce}{\begin{center}}
\newcommand{\ece}{\end{center}}
\newcommand{\beq}{\begin{equation}}
\newcommand{\eeq}{\end{equation}}
\newcommand{\bea}{\vspace{0.25cm}\begin{eqnarray}}
\newcommand{\eea}{\end{eqnarray}}

\newcommand{\ba}{\begin{array}}
\newcommand{\ea}{\end{array}}

\newcommand{\r}{\mbox{{\boldmath
$\rho$}}}

\newcommand{\qb}{\mbox{{\bf
q}}}

\def\lsim{\mathrel{\rlap{\lower4pt\hbox{\hskip1pt$\sim$}}
    \raise1pt\hbox{$<$}}}         
\def\gsim{\mathrel{\rlap{\lower4pt\hbox{\hskip1pt$\sim$}}
    \raise1pt\hbox{$>$}}}         

    \def\beq{\begin{equation}}
    \def\endeq{\end{equation}}
    \def\bea{\begin{eqnarray}}
    \def\arr{\begin{eqnarray}}
    \def\eea{\end{eqnarray}}

\def\q2{$Q^{2}$}
\def\s2{2$S$}

\begin{document}
\thispagestyle{empty}
\vspace*{-2cm}
\begin{flushleft}  
\hspace{7.3cm}Talk given at 33rd Rencontres de Moriond\\
\hspace{7.3cm}"QCD and High Energy Hadronic Interactions",\\
\hspace{7.3cm}Les Arcs, France, March 21--28, 1998\\
\end{flushleft}
\bigskip

\begin{center}

  {\Large\bf
  Quark energy loss in an expanding quark-gluon
  plasma\\
  \vspace{1.0cm}
  }
\medskip
  {\large
  B.G. Zakharov
  \medskip
  \\
  }
  {\it
  Max--Planck Institut f\"ur Kernphysik, Postfach
  103980\\
  69029 Heidelberg,
  Germany\medskip\\
  L. D. Landau Institute for Theoretical
  Physics,
  GSP-1, 117940,\\ ul. Kosygina 2, 117334 Moscow,
  Russia
  \vspace{1.7cm}\\
  }
  
  {\bf
  Abstract}
\end{center}
{
\baselineskip=9pt
We study the quark energy loss in an expanding quark-gluon
plasma.
The expanding plasma produced in high energy $AA$-collision is described by
Bjorken's
model. The dependence of the energy loss on the infrared cutoff
for
the radiated gluons, on the quark mass, and on the initial conditions
of
QCD plasma is
investigated.
}
\pagebreak
\newpage

Study of the induced gluon radiation from a fast quark
in a
quark-gluon plasma (QGP) is of great
importance
in connection with the forthcoming
experiments
on high energy $AA$-collisions at the RHIC and
LHC.
It is expected that the energy loss
of
high-$p_{\perp}$ jets produced at the initial stage
of
$AA$-collision
may be an important potential probe for formation of
QGP.$^{1,2)}$
In the recent works$^{3,4)}$ the quark energy
loss,
$\Delta E_{q}$, was estimated
for
a homogeneous
QGP. Here
we report on the evaluation of
$\Delta
E_{q}$ in expanding QGP in the light-cone path
integral
approach to the induced radiation.$^{5)}$
As in previous works, QGP is modelled by a system
of static
Debye screened scattering
centers.$^{1)}$

We consider a fast quark produced in the central rapidity region with a
velocity perpendicular to the axis of
$AA$-collision.
We choose the $z$ axis along the initial quark
velocity,
and the quark production point is assumed to be at
$z=0$.
Then, the
distance
passed by the quark in QGP, $L=z$, is close to
the
expansion time,
$\tau$.
For the $\tau(z)$-dependence of the temperature of
QGP
we use prediction of Bjorken's
model$^{6)}$
$T\tau^{1/3}=T_{0}\tau_{0}^{1/3}$.

The probability of radiation of a
gluon
with the fractional longitudinal momentum $x$ from a fast
quark
produced at $z=0$ is given
by$^{5)}$
\beq
\frac{d P}{d
x}=2\mbox{Re}\!
\int\limits_{0}^{\infty}\! d
z_{1}\!
\int\limits_{z_{1}}^{\infty}d
z_{2}
\exp\left[\frac{i(z_{1}-z_{2})}{L_{f}}\right]
g(z_{1},z_{2},x)\left[{K}(0,z_{2}|0,z_{1})
-{K}_{v}(0,z_{2}|0,z_{1})\right]\,.
\label{eq:10}
\eeq
Here $K$
is
the Green's
function
for the Hamiltonian (acting in the transverse
plane)
\beq
{H}=\frac{{\qb}^{2}}{2\mu(x)}+v(\r,z)\,,
\label{eq:20}
\eeq
\beq
v(\r,z)=-i\frac{n(z)\sigma_{3}(\rho,x,z)}{2}\,,
\label{eq:30}
\eeq
and $K_{v}$ is the Green's
function
for the Hamiltonian (\ref{eq:20}) with
$v(\r,z)=0$.
In
(\ref{eq:20}) the Schr\"odinger mass is
$\mu(x)=E_{q}x(1-x)$,
$L_{f}={2E_{q}x(1-x)}/{[m_{q}^{2}x^{2}+m_{g}^{2}(1-x)]}\,\,$
is the gluon formation length,
here
$m_{q}$ is the quark mass, and $m_{g}$ is the mass of the
radiated
gluon. We introduce the gluon mass to remove the
contribution of the unphysical long-wave
gluon
excitations.
In (\ref{eq:30}) $n(z)$ is the number density of QGP, and $\sigma_{3}$
is
the cross section of interaction of color singlet $q\bar{q}g$ system
with color center.  Summation over
triplet (quark) and octet (gluon) color states is
implicit in (\ref{eq:30}).
The $z$-dependence of $\sigma_{3}$ is connected with
the one of the Debye screening mass.

The vertex factor $g(z_{1},z_{2},x)$, entering
(\ref{eq:10}),
reads (we neglect the
spin-flip
$q\rightarrow qg$ transitions,
which
give a small contribution to the quark energy
loss)
\beq
g(z_{1},z_{2},x)=
\frac{\alpha_{s}[4-4x+2x^{2}]}{3x}\,
\frac{\qb(z_{2})\cdot\qb(z_{1})}{\mu^{2}(x)}\,.
\label{eq:40}
\eeq

The Hamiltonian (\ref{eq:10}) describes evolution of the
light-cone
wave function of a fictitious $q\bar{q}g$
system.$^{5)}$
Using relations between the transverse Green's
functions
and the light-cone functions for the
transition
$q\rightarrow qg$ in vacuum and inside
medium$^{7)}$
the radiation rate (\ref{eq:10}) after some algebra can
be
represented in another
form
\beq
\frac{d P}{d
x}=
\int\limits_{0}^{\infty}\! d z\,
n(z)
\frac{d
\sigma_{eff}^{BH}(x,z)}{dx}\,,
\label{eq:50}
\eeq
\beq
\frac{d
\sigma_{eff}^{BH}(x,z)}{dx}=\mbox{Re}
\int d\r\,
\psi^{*}(\r,x)\sigma_{3}(\rho,x,z)\psi(\r,x,z)\,,
\label{eq:60}
\eeq
where $\psi(\r,x)$ is the light-cone wave function
for
the $q\rightarrow qg$ transition in vacuum,
and
$\psi(\r,x,z)$ is the
medium-modified light-cone wave function for
$q\rightarrow qg$ transition in medium at the
longitudinal coordinate $z$. In the low density limit
and
$z\rightarrow
\infty$ $\psi(\r,x,z)$ becomes close to $\psi(\r,x)$,
and
(\ref{eq:60}) goes over into formula for
the
Bethe-Heitler cross section for a quark incident on an
isolated
center,
$d\sigma^{BH}(x)/dx$.$^{8)}$
At $z\rightarrow
0$
the ratio
$d\sigma^{BH}_{eff}(x,z)/dx/d\sigma^{BH}(x)/dx$
vanishes.$^{4,8)}$
This is a direct consequence of the decrease at small $z$ of
the transverse size of the $qg$ Fock component of the
quark
produced at
$z=0$.
Note
that
this effect is responsible for the
$L^{2}$-dependence
of $\Delta E_{q}$ at $E_{q}\rightarrow
\infty$
for a homogeneous medium.$^{3,4)}$

The three-body cross section entering
the
imaginary potential (\ref{eq:30}) can be written
as$^{9)}$
\beq
\sigma_{3}(\rho,x,z)=\frac{9}{8}[\sigma_{2}(\rho,z)
+\sigma_{2}((1-x)\rho,z)]-\frac{1}{8}\sigma_{2}(x\rho,z)\,,
\label{eq:70}
\eeq
where $\sigma_{2}(\rho,z)$ is the dipole cross section
of
interaction with color center of color singlet $q\bar{q}$
system.
The latter can be written
as
$
\sigma_{2}(\rho,z)=C_{2}(\rho,z)\rho^{2}\,,
$
where $C_{2}(\rho,z)$ has a smooth (logarithmic) dependence
on
$\rho$ at small
$\rho$.
As in Ref. 4 we
approximate
$C_{2}(\rho,z)$ by its value at $\rho\approx
1/m_{g}$.
Then the Hamiltonian (\ref{eq:20}) takes the oscillator
form
with the $z$-dependent
frequency
\beq
\Omega(z)=
\frac{(1-i)}{\sqrt{2}}
\left(\frac{n(z)C_{3}(x,z)}{E_{q}x(1-x)}\right)^{1/2}\,,
\label{eq:80}
\eeq
where
$
C_{3}(x,z)=\frac{1}{8}\left\{9[1+(1-x)^{2}]-x^{2}\right\}
C_{2}(1/m_{g},z)\,.
$
The coefficient $C_{2}$ was calculated in the double
gluon
approximation. In numerical calculations we evaluated the
Green's
function $K$ using the approach previously developed
in analysis of the vector meson
photoproduction.$^{10)}$

It follows from (\ref{eq:80}) that even for a thermolized QGP the oscillator
frequency is dominated by the gluon
contribution. 
There is every indication$^{11,12)}$ that
for
the RICH and LHC the hot QCD medium
produced
in $AA$-collisions at $\tau \sim \tau_{0}\sim 0.1$
fm
will be thermolized gluon plasma to a first
approximation, and for quarks
the chemical equilibration
is not reached during the expansion of
QGP.
For this reason
we
take for the gluon fugacity $\lambda_{g}=1$. To take
into
account the suppression of quarks we use the
value
$\lambda_{q}=1/3$ for the quark
fugacity.
Numerical calculations were carried out with
$\alpha_{s}=1/3$.
For $\tau_{0}$ we use the value 0.1
fm.$^{11,12)}$
At $z\le \tau_{0}$ we take
$\Omega(z)=\Omega(\tau_{0})$.
We have evaluated $\Delta E_{q}$ for
$T_{0}=1100$
expected for central Pb-Pb collisions at the
LHC.$^{11)}$
To study the dependence of the quark energy loss on $T_{0}$
we
give also the results for $T_{0}=700$
MeV.
The numerical predictions for $\Delta E_{q}$ as a function of
$E_{q}$
for $L=3,\,6,\,9$ fm are shown in
Fig.~1.
The results were obtained for $m_{q}=0.2$ GeV. As for
a homogeneous QGP$^{4)}$, our predictions have a
weak
dependence on $m_{q}$.
To study the infrared sensitivity of $\Delta E_{q}$
we present in Fig.~1 the results
for
$m_{g}=0.75$ and 0.375 GeV. These values are of the order
of
the Debye screening mass for QGP in the region $\tau\gsim 2-3$
fm,
which, as our numerical calculations show, dominates the quark
energy
loss. For this reason the value of $m_{g}$ within the above
range
seems to be a plausible estimate for the infrared cutoff in
the
considered
problem.
Note that in the
limit
$E_{q}\rightarrow
\infty$
$\Delta E_{q}$ has only a smooth (logarithmic)
$m_{g}$-dependence
connected with $\rho$-dependence of the coefficient
$C_{2}$.$^{4)}$
Our numerical calculations really
demonstrate
that for $E_{q}\gsim 10$ GeV $\Delta E_{q}$ is not very
sensitive
to
$m_{g}$.
Fig.~1 shows that $\Delta E_{q}$
grows
by a factor $\sim 2-3$ as $E_{q}$ increases from $\sim 10$
to $\sim 100$ GeV. The predictions for $L=6$ and $L=9$
fm
can be regarded as a plausible estimates for $\Delta
E_{q}$
in central collisions of heavy nuclei. We have checked
that
the effect of the mixed phase (with $T_{c}\approx 150$ MeV)
and
the hadronic phase turns
out
to be relatively
small.
Comparison of the results for $T_{0}=1100$ and $T_{0}=700$
MeV
demonstrates that the $T_{0}$-dependence of $\Delta
E_{q}$
is not very strong.
This is connected with
an increase of 
Landau-Pomeranchuk-Migdal
suppression and  a decrease of the coefficient $C_{2}$
for a high density QGP.

To study the sensitivity of $\Delta E_{q}$ to dynamics
of
QGP at small $\tau$ we have also carried out the calculations
taking $\Omega(z\lsim a)=\Omega(a)$ with $a\sim 1-2$
fm.
The results for these versions do not differ
strongly
from the ones in Fig.~1. This demonstrates
that
$\Delta E_{q}$ is insensitive to the dynamics of QGP at $\tau\lsim
2$
fm. This fact is a consequence of the vanishing of the
ratio
$d\sigma^{BH}_{eff}(x,z)/dx/d\sigma^{BH}(x)/dx$ at small
$z$.

To illustrate the quark mass dependence of $\Delta E_{q}$
in
Fig.~2 we compare the results for $\Delta E_{q}$
for
$c$-quark ($m_{q}=1.5$ GeV) with the ones for light
quarks
($m_{q}=0.2$ GeV). As one can see the dependence of $\Delta E_{q}$
on
$m_{q}$ becomes weak at $E_{q}\gsim 50-100$
GeV.

In Fig.~3 we present the gluon
spectra
for $m_{q}=0.2$ and 1.5 GeV at $E_{q}=$50, 100 and 200
GeV
for $T_{0}=1100$ MeV. We also show in this figure the
Bethe-Heitler
spectra. As one can see the radiation rate
is
strongly suppressed by the medium and finite-size
effects
for the gluon momenta $k\gsim 10$ GeV (excluding a narrow
region
near $k\approx E_{q}$). For the heavy quark the radiation
rate
at $k\approx E_{q}$ is suppressed as compared to the light
quark.
However, in the soft region the spectra for the light and heavy
quarks
are close to each
other.


Our predictions
for
the absolute normalization of the energy loss  must
be
regarded as rough estimates with uncertainties of a factor $\sim
2$.
Nevertheless, the obtained rather large values of $\Delta
E_{q}$
show that jet quenching may be an important probe for
the
formation of QGP in
$AA$-collisions.
The Monte-Carlo analysis of jet
quenching$^{2)}$
with $\Delta E_{q}$ close to the estimates of the present
work
demonstrates that the induced radiation must
considerably
modify the charged particle spectra. The induced
gluon radiation
can also lead to other interesting experimental
consequences.
For instance, fluctuations of the quark energy loss
will
generate additional transverse momentum (defined with respect to
the
$AA$-collision axis) in production of $c\bar{c}$ and
$b\bar{b}$
pairs. In the case of $g\rightarrow
gg$
transitions the production of the $gg$ state in the color
decuplet
state can increase the cross section for production
of
baryon-antibaryon pairs through the
mechanism
analogous to the one previously discussed$^{13)}$ in connection
with
$B\bar{B}$ annihilation at high
energies.

This work was partially supported by the INTAS
grants
93-239ext and
96-0597.


\newpage
\begin{center}
{\Large \bf Figures}
\end{center}
\begin{figure}[h]
\begin{center}
\epsfig{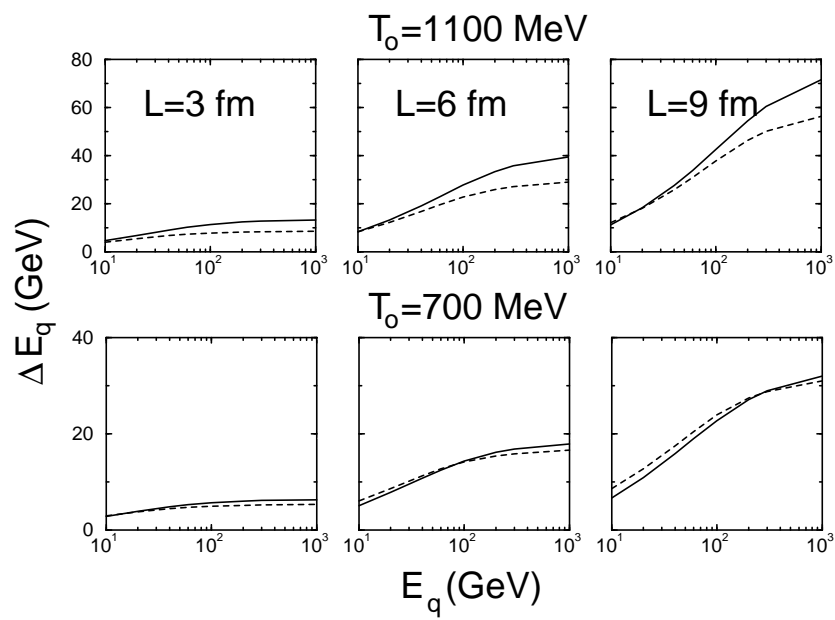}
\end{center}
\vspace{-.3cm}
\caption[.]{
The quark energy loss as a function of the quark energy
for $m_{q}=0.2$ GeV, $m_{g}=0.75$ GeV (solid line) and $m_{g}=0.375$ GeV
(dashed line).
}  
\end{figure}
\begin{figure}[h]
\begin{center}
\epsfig{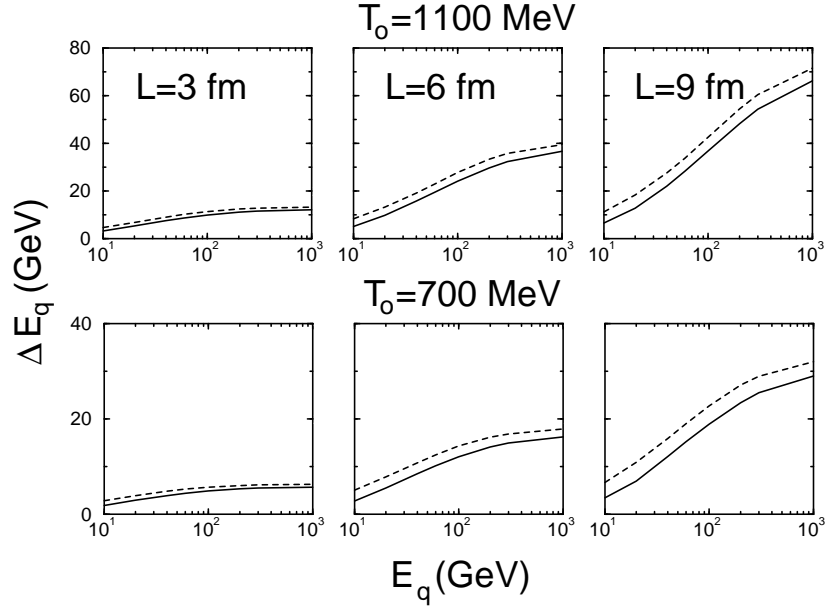}
\end{center}
\vspace{-.3cm}
\caption[.]{
The quark energy loss as a function of the quark energy
for $m_{g}=0.75$ GeV, $m_{q}=1.5$ GeV (solid line) and $m_{q}=0.2$ GeV
(dashed line).
}  
\end{figure}
\begin{figure}[h]
\begin{center}
\epsfig{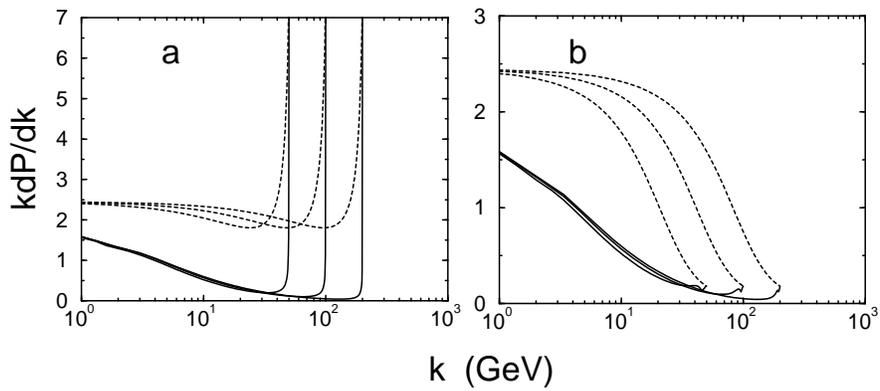}
\end{center}
\vspace{-.3cm}
\caption[.]{The gluon spectrum (solid line) as a function of the gluon
momentum
at $E_{q}=$50, 100 and 200 GeV for $m_{q}=0.2$ (a) and 1.5 (b) GeV.
In both the cases $m_{g}=0.75$ GeV, and $T_{0}=1100$ MeV.
The dashed line shows the Bethe-Heitler spectrum.  
}  
\end{figure}

\end{document}